# Space Qualifying Silicon Photonic Modulators and Circuits

Dun Mao[1,2†], Lorry Chang[1†], Hwaseob Lee[1], Anthony W. Yu[3], Bennett A. Maruca[4], Kaleem Ullah[1], William H. Matthaeus[4], Michael A. Krainak[3], Po Dong[2*], and Tingyi Gu[1*]

[1] Department of Electrical and Computer Engineering, University of Delaware, Newark, DE 19716

[2] II-VI Incorporated, 48800 Milmont Drive, Milmont, CA 94538

[3] NASA Goddard Space Flight Center, Lasers & Electro-Optics Branch, Greenbelt, MD 20771

[4] Department of Physics and Astronomy, University of Delaware, DE 19716

[†] Authors contributed equally

* Email: tingyigu@udel.edu, po.dong@ii-vi.com

**Abstract:** Reducing the form factor while retaining the radiation hardness and performance matrix is the goal of avionics. While a compromise between a transistor's size and its radiation hardness has reached consensus in micro-electronics, the size-performance balance for their optical counterparts has not been quested but eventually will limit the spaceborne photonic instruments' capacity to weight ratio. Here we performed the first space experiments of photonic integrated circuits (PICs), revealing the critical roles of energetic charged particles. The year-long cosmic radiation does not change carrier mobility but reduces free carrier lifetime, resulting in unchanged electro-optic modulation efficiency and well-expanded optoelectronic bandwidth. The diversity and statistics of the tested PIC modulator indicate the minimal requirement of shielding for PIC transmitters with small footprint modulators and complexed routing waveguides, towards light-weight space terminals for terabits communications and inter-satellite ranging.

**Teaser:** Nanoscale optoelectronic devices and circuits may transcend the limits for future space instrumentations.

## Introduction

Optical communication with terabit capacity is demanded by key applications ranging from near earth to deep space communications, and from human-space exploration to astrophysics experiments [1-3]. Photonic integrated circuits (PICs) with extreme energy efficiency, bandwidth [4-7], and system-on-chip capacities [8-9] have emerged as promising candidates for space-borne optical communication instruments and adaptive signal processors [10-14]. Silicon (Si) PICs are foundry manufacturable, and the information exchange and broadcasting are carried out by optical wave-guides (WGs), which drastically expands the throughputs compared to copper interconnect, and naturally less sensitive to electromagnetic interferences, charge-related total ionizing dose (TID) and transient single-event effects. The displacement damage (DDD) of the particle radiation-induced defects is the primary concern for bulk optoelectronic components. Understanding the PIC's response to high-energy particle radiation not only facilitates this nanophotonic technology to be infused into future flight missions involving electronic circuits [15] and fiber-based technologies [16]. Such knowledge also provides scientific insights toward the explorations of nanoscale photonic modulation of energetic particle beams, particle-photon entanglement, and particle quantum optics [17-18].

Cosmic radiation covers a wide set of radiation sources with a broad range of energy spectra. Space radiation is composed of two primary groups: cosmic rays of electromagnetic wave packets, and radiation particles (protons, electrons, and heavy ions) reaching extremely high kinetic energy [16]. Ground radiation tests of Si PICs have been focusing on single frequency gamma ($10^6$ eV) [19-22] and X-ray ($10^4$ eV) exposure of passive devices [19, 23, 24-25], with dosage equivalent to $10^3$ years of exposure on LEO. The high dosage of such electromagnetic radiation results in primary TIDs in silicon-on-insulator (SOI) and reduces the tuning efficiency [23, 25-26]. Particle radiation (especially protons and heavy ions) is considered the primary cause of degradation for bulk opto-electronics in space [27-29], but no study has been found on the particle radiation impacts on the nanoscale-doped photonic circuits. Brasch et al. combine four proton sources for reproducing the particle radiation energy spectra close to LEO and exposed passive $SiN_x$ PICs [30], but the high energy portion of the cosmic radiation particles is still missing. We know also the exposed material is undoped $SiN_x$ PICs, rather than the doped Si that we study here. Such high-energy proton exposure can lead to cluster defects [31] and cannot be shielded even with a 10mm Aluminum (Al) sheet [30].

When the high-speed cosmic particles incident on the device's top surface, their kinetic energy attenuates with their penetration depth. Cascaded atomic structural damage is expected along its trajectory as the energetic particle collides with adjacent and secondary atoms. Monte Carlo simulations capture that such nuclei displacements scale with the particle mass but are inversely correlated with the particle velocity [32-33]. In bulk optoelectronic components, the speed-attenuated orbital protons create nuclei vacancies and displaced atoms in the active layer embedded at least a few μm below the top surface, leading to device degradation [34-35]. In optical fibers, increased insertion loss and change in refractive index were observed after cosmic radiation exposure [36-37]. However, the high energy proton radiation impacts on nanoscale optoelectronic circuits are fundamentally different. Minimal nuclei displacement is expected given orbital protons' high kinetic energy and limited attenuations when they reach the top device layer of SOI, but little is known about the optoelectronic device response to the charged particle interactions with the bounded electrons. Especially, the charged particles' inelastic scattering with bounded electrons takes place at much higher kinetic energy (up to 100MeV) [38] compared to nuclei (k-MeV).

**Results**

Leveraging the manufacturing capability of Si photonic foundries, we performed the first space test of subwavelength photonic devices by exposing arrays of Si PIC devices and circuits on LEO for nearly 11 months. The PICs from different foundries were mounted on the Materials International Space Station Experiment-Flight Facility (MISSE-FF) located near the exterior of the International Space Station (ISS) (Fig. 1A). The radiation dosage is expected to approach the amount received by the optical instruments in CubeSat missions [3]. Si photonic modulators with μm and sub-cm size (Fig. 1B-C) were prepared to compare their direct current, continuous wave response, high-speed optoelectronic, and nonlinear optic responses before and after LEO exposure. Limited charge scattering centers are created, given the unchanged carrier mobility, series resistance, and electro-optic tuning efficiency. The significantly reduced free carrier lifetime ($\tau_{rec}$), reflected as the expanded optoelectronic bandwidth and reduced thermal nonlinearities, is attributed to the high energy protons created dangling bonds in the top Si layer on SOI. Through characterizing arrays of doped waveguides with different lengths, we observe the evidence of the heavy ion degradation and damaging trajectories along SOI.

We prepared one passive die with arrays of high-$Q$ microring resonators (MRRs) from the American Institute of Manufacturing of Integrated Photonics (AIM Photonics), and two active dies with Mach Zehnder Modulators (MZMs) and Microring Modulators (MRMs) from the Institute of Microelectronics, Singapore (IME) for LEO exposure [39]. Other dies from the same multi-project-wafer (MPW) runs were kept in the lab as control samples (methods). The samples sent to space are wrapped in paper sleeves and protected under a thin Al sheet (0.110 inch or 2.794mm) against micrometeorites. Note that the Al shielding effect is very limited for γ-rays and highly energetic particles (>$10^7$eV/particle) that are prevalent on LEO [30]. Such a thin Al sheet can greatly attenuate most of the low energy radiation (<1MeV/particle) in the ground test. The suitcase enclosed those samples unfolded in space, facing towards the direction of the flight (RAM) on ISS. The orbital temperature cycles are stabilized between -30 and +40°C. The exposure lasted 325 days 10 hours and 11 minutes (approximately 6,700 orbits) (methods). The daily cosmic radiation dosage was recorded by the Si-based dosimeters installed on ISS, composed of the galactic cosmic radiation (GCR) and south Atlantic anomalous (RAA) (method). The interplanetary particles of GCR are mostly protons (~80%) and alpha particles (10%), with energies ranging from $10^7$ to $10^{11}$ eV per particle [27]. Specifically, the GCR includes heavy ions of more than 50MeV, created during their path through the interstellar gas. The integral of the daily dosage (GCR and RAA) results in the accumulated total radiation dosage of 148.5mGy or 14.85 rad (Fig. 1D).

The LEO-exposed MPW dies include arrays of the carrier depletion-based Si photonic modulators, including 10 µm radius MRMs (Fig. 1B) and MZMs with the active arm length of sub-cm (Fig. 1C). During the device characterizations, the light was coupled on and off the chip through lensed fiber and edge coupler. The bias and RF electronic signals are sent to the chip through high-speed GSG probes (shown in Fig. 1C) (Materials and Methods). The Si single-mode WGs define the optical paths for making the interferometric circuits (marked as blue lines in Fig. 1B-C). Different from most of the ground test reports, our experiments show that the electro-optic tuning efficiency remains invariant after radiation exposure for both MRMs (Fig. 1E) and MZMs (Fig. 1F). The models for fitting the transmission spectra of MRMs and MZMs are detailed in the Supplementary Materials (SM) S1. The extinction ratio (ER) drops from near 24 dB to 14 dB for MRMs (Fig. 1G) and from 26 dB to 17.5 dB for MZMs (Fig. 1H). By fitting the coupled-mode-theory (CMT) model to the transmission spectra (equation S-2), the intrinsic quality factor ($Q_{in}$) for the MRM is ex-tracted for pre- and post-flight devices (drops from 36k to 25k at zero bias), indicating 30% increase in propagation loss $\alpha$. Reduced sensitivity of ER and extinction coefficient ($\Delta k$) to reverse bias is observed in both MZMs and MRMs (Fig. 1H).

The propagation loss of active WGs is characterized by measuring the arm length-dependent fiber-to-fiber loss of MZMs (Fig. 2A). The incremental propagation loss induced ER reduction was mapped across an MPW die with more than 40 devices (Fig. 2B). Analytical model provides the device dependent refractive index ($\Delta n$) and $\Delta k$ change (Fig. 2C-E). The extracted $\Delta k$ of $10^{-3}$ aligns with the directly measured incremental propagation loss of 20dB/cm (Fig. S1A). We also measured passive reference WGs with lengths up to 7 mm. No incremental propagation loss was observed after LEO exposure (Fig. S1B). Note that the post-flight MZMs in the shaded area in Fig. 2B failed as one of the cm-long active arms suffered from an additional loss of more than 10 dB. This WG damage, which is only observed in large-footprint MZMs, aligns with the characteristic behavior of heavy ions (SM S3). On the same die, arrays of MRRs with varying ring-WG coupling gap sizes and coupling quality factor ($Q_c$) were included. The ER of MRR is maximized near critical coupling, where the $Q_{in}$ is close to $Q_c$. Measured transmission spectra for a set of MRMs indicate the critical coupled resonator moved from the one with a larger gap size to the bus WG (higher $Q_c$) to the one with a smaller gap (Fig. 2F). The $Q_{in}$ reduction is verified through CMT fittings of the MRMs arrays (Fig. 2G-H).

The impact of the LEO exposure on micro-electronic properties is characterized across tens of *p-i-n* junctions in MZM and *p-n* junctions in MRM. The standard deviation of the electronic characterizations among the optoelectronic devices is included as the error bars in Fig. 3A-D. No significant change in capacitance or series resistance of the junctions was observed (Fig. 3 A,C), where series resistance is inversely related to carrier mobility [40]. Increased ideality factor (average value from 1.37 to 1.43) and reduced reverse saturation current ($I_0 \propto \tau_{rec}^{-1/2}$) are identified by fitting the IV curves of *p-n* and *p-i-n* junctions (SM S4) (inset of Fig. 3B) [40]. The high-speed optoelectronic spectra of that carrier depletion-based Si photonic MZMs and MRMs were carried out using identical experimental apparatus (same power level of optical carrier, and RF signal) for pre- and post-flight samples (Method and Supporting Information file S5) (Fig. 3E-H). Within >30 measurement datasets per device for 12 devices (3 pairs of MZMs and 3 pairs of MRMs), it is conclusive that S21 spectra show well-expanded 3dB optoelectronic bandwidths across multiple modulators with different device footprints. The faster modulator response, along with the increased ideality factor and reverse saturation current, indicates the reduced $\tau_{rec}$. Note that the photon lifetime in MRM is around 5ps given the $Q_t$, which is less than 5% compared to the electronic limited response time.

The impact of $\tau_{rec}$ reduction is also verified through steady-state nonlinear optic measurement in passive MRRs (from AIM) and active MRMs (from IME). Nonlinear coupled mode equations derive that the nonlinear absorption generated free carrier density linearly increases with the $\tau_{rec}$ at steady state excitations, and thus the photo-thermal effect induced resonance shift [41-42] (SM S6). Note that this approximation does not hold at higher optical excitation power well beyond the nonlinear threshold. Comparing the pre-flight and post-flight MRR with the same $Q_t$ (~24k), a more significant resonance shift is observed in the pre-flight sample at the same power levels (Fig. 4A). The power-dependent photothermal resonance shift shows consistent results (Fig. 4B). The reduced $\tau_{rec}$ and nonlinear photothermal response are also verified in the MRMs (Fig. 4 inset), where two MRR with similar $Q_t$ (~22k) and same *p-n* junction designs were selected for comparison. Since the TID effects are usually recovered after thermal activation, we performed the post-flight annealing and examined the $\tau_{rec}$ by tracking the nonlinear resonance shift. The nonlinear optical response does not change after heating up to 300ºC for an hour (Fig. S5). The reduced $\tau_{rec}$ is attributed to the dangling bonds formed by the high energy protons swift through the 220nm thick Si layer on SOI (Fig. 4C) [38], while the proton with low kinetic energy (slowed down cosmic radiation in bulk device) leads to more extensive damage of nuclei displacement (or DDD) (Fig. 4D). Those displaced nuclei and vacancies act as carrier scattering centers and reduce the carrier mobility. Given the unchanged series resistance and electro-optic tuning efficiency, the carrier mobility remains the same after LEO exposure.

**Discussion**

Despite numerous proposals of system-on-chip integration of optical modules for avionics, limited literature is found on space radiation impacts on nanoscale devices, which are critical in propelling nanotechnology for space instrumentation. In addition to the unknown radiation spectra, the device-to-device variations are another challenge blurring the underlying physical mechanism. In this work, we advance the understanding of these issues by sending foundry-manufactured dies with hundreds of passive and active devices beyond 2,000 km altitude. Those nanoscale optoelectronic devices and circuits are characterized by the electronic, optical, high-speed optoelectronic, and nonlinear optics aspects. Studies in bulk semiconductor components indicate the charged energetic particles recombine or migrate and form stable defects in the Si lattice structure, resulting in reduced carrier mobility and optoelectronic efficiency [43-44]. No study has been found on the effects of particle radiation on doped PICs. Interestingly, the reduced $\tau_{rec}$ in the Si device layer leads to optoelectronic bandwidth improvement of MZMs and MRMs. In contrast, the sub-cm long active long arm with *p-n* junctions shows a few dB excess loss, which results in reduced ER in both MZMs and MRMs. Nevertheless, after long-term exposure, the ER stays beyond 10 dB, providing sufficient modulator amplitude. Remarkably, the tuning efficiency of both types of modulators remains unchanged, unlike many ground radiation exposure results on PIC. The decreased electro-optic tuning efficiency reported in the ground test is induced by the high dosage X-ray or γ-ray exposure (typically Mrad), whose mechanism is completely different from the space radiation. Also, we observed that four adjacent long WGs with *p-i-n* junctions suffered from very high insertion loss after LEO exposure, likely attributed to the accidental deposition of heavy ions and formation of nanoscale cluster defects. The behavior of heavy ions (mostly coming from galactic cosmic rays) aligns with the characteristic of low possibility but disruptive damage to the Si nanowire, including low flux ~4 particles cm$^{-2}$s$^{-1}$, high atomic number, and high energy. The deposition and reflection of a single heavy ion on Si WG arrays lead to the disconnection of multiple cm-long active WGs on the same row.

We summarized the LEO exposure impacts on optical properties (Table S-I) and optoelectronic response (Table S-II) of the Si photonic devices, after evaluating nine aspects (listed in Table S-IV) of the nanophotonic device responses over 100 devices. The error bars are included the plots and histograms for refractive index change (Fig. 2D, E), capacitance (inset of Fig. 3A), ideality factors (inset of Fig. 3B), RC constant (Fig. 3C), $\tau_{rec}$ (Fig. 3D) and doping dependent propagation loss (Fig. S1). Only parameters with pre- and post-flight contrast significantly larger than standard deviations are counted towards LEO exposure impacts. Both MRMs and MZMs exhibit degradation of ER, but electro-optic resonance tuning efficiency remains invariant. The unchanged tuning efficiency and series resistance indicate the carrier mobility remains the same after LEO exposure. The change of ideality factors, reverse saturation current and the high-speed optoelectronic response of the embedded *p-n* and *p-i-n* junctions unanimously reflect a nearly half reduction of $\tau_{rec}$ (from 500 ps to 226ps), aligning with the reduced nonlinear photothermal response in MRRs in both passive and active designs. The $\tau_{rec}$ reduction apparently originates from the high kinetic energy protons that form dangling bonds in the Si lattice in SOI. Reduced $\tau_{rec}$ increases ideality factors and expands electro-optic bandwidth. While the propagation loss is unchanged in passive waveguides, the enhanced propagation loss in the doped Si waveguides (+20dB/cm) might be attributed to the defects formed by the charged radiation (protons and alpha particles). In addition, accidental disruptive damage on a few MZM with cm long arms is attributed to the cluster defects triggered by the long waveguide interaction with discrete events of heavy ions. Micro-Raman spectra of the Si transverse optic photon (SM S2) suggest the crystal symmetry in passive/undoped Si waveguides changed after LEO exposure, while the doped waveguides spectra are unchanged.

It is noted that given the timing and expense of the experiments, this very first space experiment focuses on the understanding of nanoscale optoelectronic device statistics with foundry-provided modulators with average performance. The device response time $\tau_e$ is determined by $\tau_{rec}$ and carrier transit time ($\tau_t(V)$): $\frac{1}{\tau_e} = \frac{1}{\tau_{rec}} + \frac{1}{\tau_{tr}(V)}$ (detailed in section S5). To clarify the impacts of the $\tau_{rec}$, we keep the device operating near zero bias during the high-speed characterizations. Reverse bias significantly improves $\tau_{tr}$ and thus $\tau_e$, but the impact of $\tau_{rec}$ reduction is less visible. It is expected that a higher-speed Si photonic modulator with advanced doping profile engineering is less sensitive to the $\tau_{rec}$ reduction [29]. Consecutive flight experiments can be designed to explore the way diverse radiation particles impact higher-speed Si photonic transceivers.

The small form factor and space radiation hard on-chip active nanophotonic instruments may provide a miniaturized system-on-chip platform for future astrophysics study, earth science observations, and space optical communications. Our research combines the specialties of astronomy and nanoscale optoelectronics, specifying the major impacts on Si photonics from the high-energy orbital particle radiation from the complex space environment. The ground tests with high-density (k-M rad) X-ray or γ-ray exposure on Si, a-Si, SiC, and $SiN_x$ PICs (Table S-III) result in totally different types of radiation damage of TID associated with surface oxidation (Table S-IV). In contrast, the LEO cosmic radiation dosage is much lower but carries extremely high energy particle radiation including protons, alpha particles, and heavy ions. The lightweight charged particles alter the electronic bonds and mid-gap defects without reducing carrier mobility. Through multi-modal characterizations, we conclude that the orbital radiation does not create carrier scattering centers but introduces carrier recombination centers, which result in unchanged driving voltage and reduced bit-error rates for PIC transceivers (Table S-IV), respectively. By avoiding large footprint modulators, those radiation-hard active nanophotonic components can build fully integrated systems with minimal shielding requirements for year-long orbital operation.

The radiation-tolerant PIC offers a versatile range of applications by supplanting conventional bulk optoelectronic instruments and part of radio-frequency transducers. This technology may foster low-cost terabit inter-satellite communications, support crewed space explorations, and enable precision astrophysics observations. Moreover, these PICs may find practical utility in LEO CubeSat remote sensing missions of surveying biodiversity, monitoring methane emissions, and assessing natural disasters. Beyond space applications, radiation-hardened PICs can also substitute electronic transducers, particularly for defense and nuclear operations, where highly energetic particle radiation is prevalent.

## Materials & Methods

Foundry-manufactured silicon photonic devices: The MRMs and MZMs were manufactured by IME through the MPW run. The lateral *p-i-n/p-n* diode configurations were defined by ion implantations: boron for *p*-type ($5\times10^{17}$ cm$^{-3}$) and phosphorus for *n*-type ($5\times10^{17}$ cm$^{-3}$). The intrinsic region is lightly *p*-doped ($10^{16}$ cm$^{-3}$). Heavily doped *p*++ and *n*++ regions ($1\times10^{20}$ cm$^{-3}$) were used to form Ohmic contact [38-39]. Vertical *vias* are patterned and etched on top of cladding oxide for the contact regions, followed by standard aluminum metallization for direct contact with the heavily doped Si regions.

The photonic structures were defined by 248 nm deep-ultraviolet photolithography on an 8-inch SOI wafer with a 220 nm device layer, followed by reactive ion etching. Two-step etching leaves a 90nm thick Si wing area for supporting the doping and contacts. A thick oxide cover layer is deposited for metal insulation. The passive Si channel WGs are 500nm wide and 220 nm thick. On each arm of MZM, the thickness of the wing is 90nm, which supports the doping areas of ridge WGs. Each arm of MZM has an effective modulation WG length that varies among devices. 42μm arm-length difference between the arms introduces a free-spectral range of 14 nm. The MRMs have a radius of 10 μm and varying gap distances to bus WGs. Different electronic designs of *p-n* and *p-i-n* junction were included for each optical design for the MRMs. The MPW runs through AIM Photonics and includes a set of MRRs with radii of 1.5, 2, 3, 5, 10, and 20 μm and varying gaps to bus WGs. The MRRs with a 20 μm radius were selected for comparison with the pre-flight sample.

Cosmic radiation dosage counting: The daily radiation dosage is recorded by Si-based radiation environment monitors (REMs) installed on the location COL1A2. Separation of the GCR/SAA dosage follows the methods provided for the REMs [45]. GCR data records all high-energy particles originating from outside of the solar system (interplanetary particles), and SAA records radiation doses from Earth's inner Van Allen radiation belts closest to Earth's surface.

Procedures for exposure on LEO: The samples are assembled in thin paper sleeves and delivered to the Johnson Space Flight Center on 07/11/2019. After assembling all the samples, NG-12 carrying all the payloads (named M12 panel) launched from the Wallops facility (11/02/2019). 9 days later, the NG-12 reached ISS and the payload was installed on MISSE-FF. On 12/03/2019, the M12 panel was opened towards RAM (direction of the flight) on ISS. M12 was closed on the day of 11/25/2020. SpaceX-21 fetched the M12 from ISS and splashed it down on 01/14/2021. The sample is then retrieved and delivered to the university lab on 03/10/2021.

High-speed optoelectronic measurement: The high-speed response was measured by the optical spectrum analysis method [46-47]. The optical carrier was coupled to the device via edge coupling. RF signal generated from a Vector Network Analyzer (VNA 0-67GHz) was sent to one port of the MZM transmission line through a high-speed ground-signal-ground (GSG) probe (cascade, 0-40GHz bandwidth). The other port of the MZM transmission lines was terminated by another GSG probe, providing 50 Ohm impedance matching. On the same setup configuration, broadband S11 spectra were collected by the VNA.

Junction characterization: The direct current-voltage traces were obtained by using an Agilent semiconductor parameter analyzer. The voltage-dependent junction capacitance was measured at the same time under 100 MHz small-signal modulation.

Numerical simulations: The effective index calculation was carried out using Ansys Lumerical MODE solutions.

**Acknowledgments** The authors acknowledge the discussions with G. Keeler, N. Wagner, and A. Christou on the experiments and the discussions with S. A., N. D., and S. T. from NASA on space applications. The authors appreciate the technical support from J. B. and I. K. from AEGIS for the flight experiment, and K. U. for the micro-Raman characterizations.

**Funding:** This work is sponsored by the National Aeronautics and Space Administration (80NSSC21M0222 and 80NSSC17K0526). The device characterizations are partially sponsored by the Air Force Office of Scientific Research (FA9550-18-1-0300) and the Defense Advanced Research Projects Agency (N660012114034).

**Author Contributions:**

Conceptualization: WHM, BAM, PD, MAK, TG

Methodology: DM, LC, HL, BAM, AWY, PD, KU

Investigation: DM, LC, HL, WHM, TG

Visualization: DM, LC, TG

Supervision: WHM, MAK, PD, TG

Writing-original draft: DM, LC

Writing-review & editing: DM, PD, WHM, TG

**Competing Interests:** The authors declare no competing interests.

**Data and Materials Availability:** All data needed to evaluate the conclusions in the paper are present in the paper and the Supplementary Materials.

**Figures:**

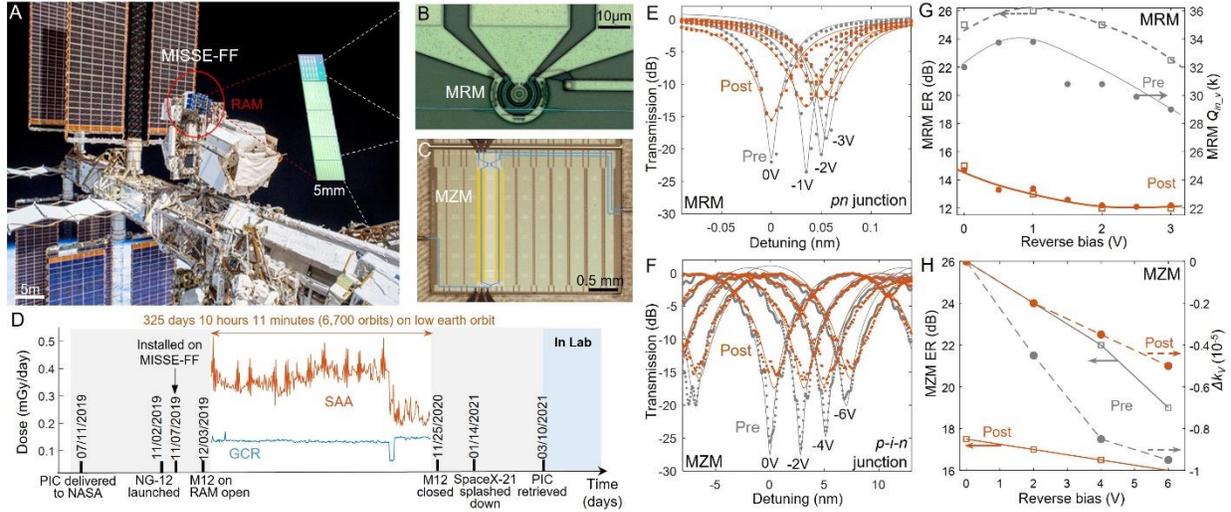

**Fig. 1. Cosmic radiation exposure on international space station** [7-8]. (**A**) Space view of the Materials International Space Station Experiment-Flight Facility (MISSE-FF) material carrier, where the PICs were mounted on the RAM direction. (**B**) Microscope image of the MRM (size of 10μm) and (**C**) MZM with sub-centimeter feature size. Blue lines: optical WG path. (**D**) A detailed timeline with the daily radiation dosage recorded by a radiation environment monitor installed on ISS. (**E**) Voltage-dependent transmission spectra of the same MRR and (**F**) MZM pre- and post-flight. Dots are measured data points and the curves are the theoretical fitting (SM S1). (**F**) The reverse bias dependent extinction ratio (ER) and extracted intrinsic quality factor $Q_{in}$ for MRM. (**G**) The reverse bias dependent ER and extinction coefficient ($\Delta k$) for MRM and (**H**) MZM. Dots are obtained by fitting the models to the experimental data. The curves are eye guides. Grey: pre-flight. Orange: post-flight.

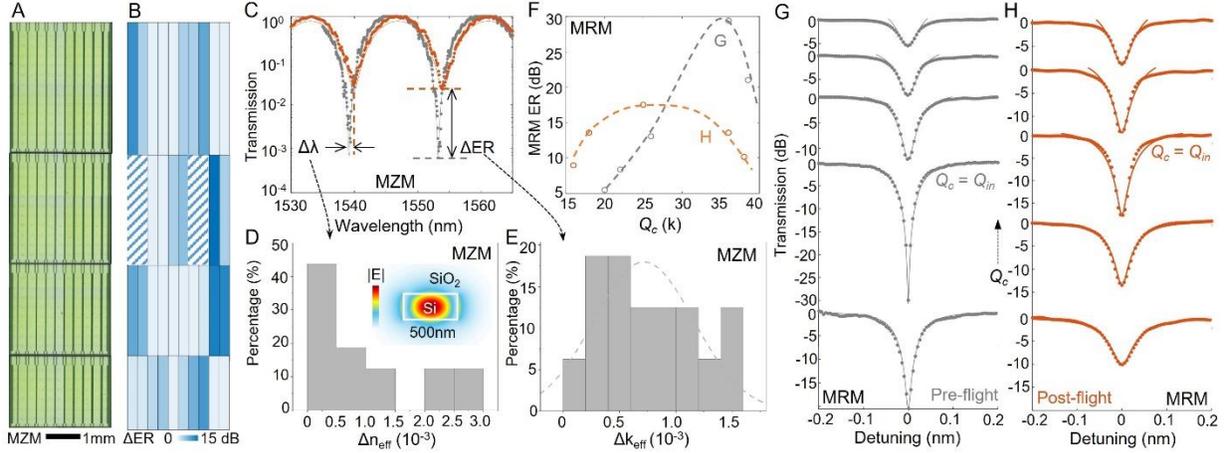

**Fig. 2. Complex refractive index contrast and modified optical responses in the interferometric silicon photonic circuits.** (**A**) Optical microscope image of a set of 40 MZMs fabricated on the same die from a multi-project wafer run. (**B**) Corresponding extinction ratio reduction ($\Delta$ER) after radiation exposure on LEO. (**C**) Representative transmission spectra of the same device on the same die pre- (grey) and post-flight (orange). The measured environmental temperature difference is less than 0.1°C. (**D**) Refractive index variation ($\Delta n_{eff}$) extracted from the resonance drift ($\Delta\lambda$), where the error bar of the temperature variation is around $10^{-4}$. (**E**) Derived extinction coefficient ($\Delta k$) for the same sets of the MZM pre- and post-flight. (**F**) The ER versus coupling quality factor ($Q_c$) of a set of MRMs with the varying gap between the resonator and bus WG. (**G**) Transmission spectra of the MRM before and (**H**) after exposure to LEO.

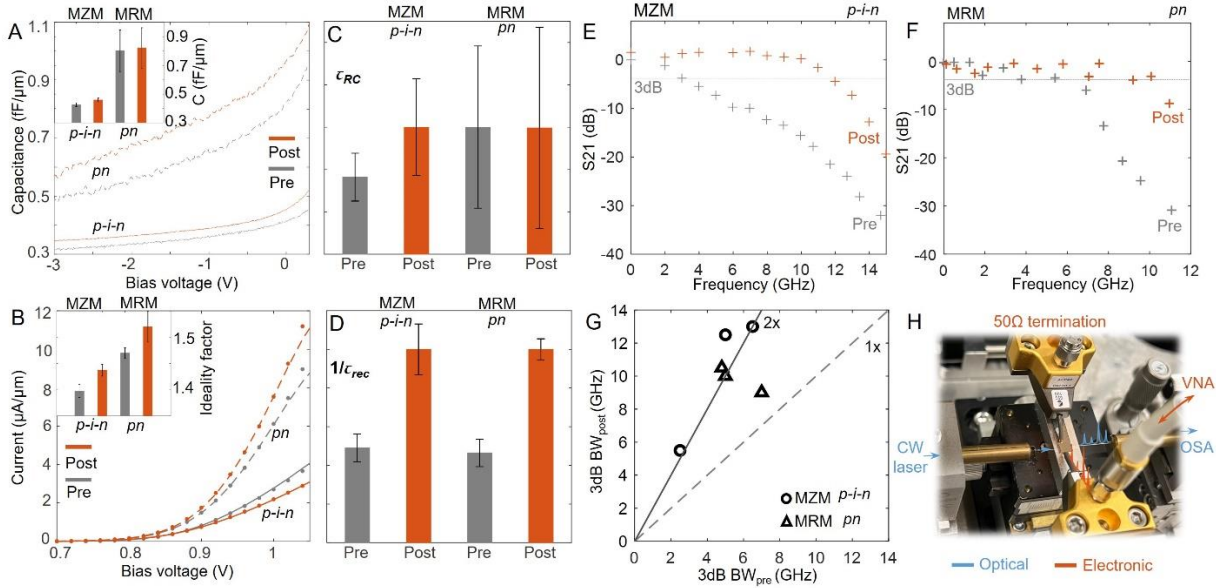

**Fig. 3. Electronic and high-speed optoelectronic characterizations of silicon photonic modulators.** (**A**) Exemplary capacitance-voltage curves and (**B**) current-voltage curves for the silicon photonic modulators with *p-n* (dashed) and *p-i-n* junctions (solid curves). Insets: statistical average of normalized capacitance (in **A**) and extracted ideality factors (in **B**) for pre- (grey) and post-exposure (orange). (**C**) Average carrier loss rate by RC constants ($\tau_{RC}$) and (**D**) by carrier lifetime ($\tau_{rec}$), extracted from (**A**) and (**B**) respectively. Error bars represent the standard deviation. (**E**) Measured electro-optic modulation response S21 of an MZM pre- (grey) and post-flight (orange). (**F**) S21 comparison for MRM. (**G**) 3dB optoelectronic bandwidth of post-flight devices versus pre-flight devices for the MZMs (circles) and MRMs (triangles). (**H**) Setup for characterizing the high-speed response of MZM.

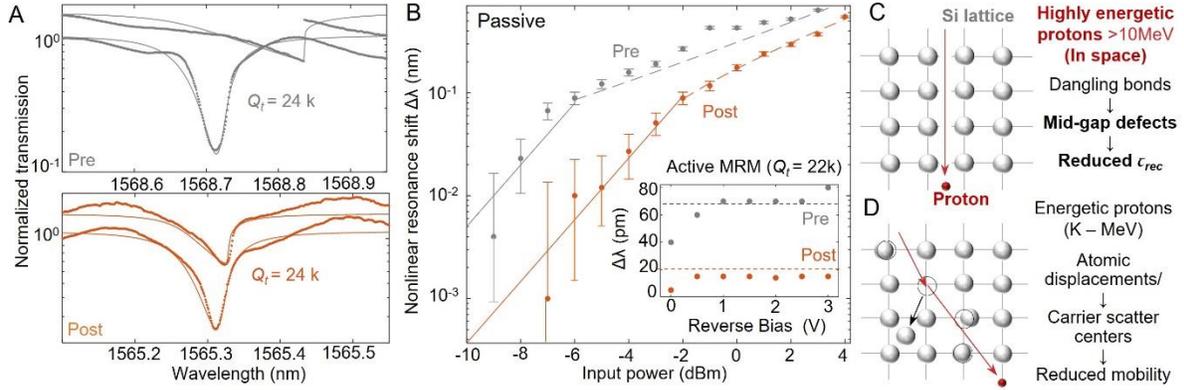

**Fig. 4. Reduced free-carrier lifetime ($\tau_{rec}$) characterized by the photo-thermal nonlinearities in the MRRs and MRMs.** (**A**) Nonlinear transmission spectra for pre-flight sample (upper panel) and post-flight sample (lower panel). The two selected MRRs have the same total quality factor ($Q_t$). Dots: experimental data. Curves: CMT fits. (**B**) The reduced nonlinear response is characterized by the input power-dependent resonance wavelength detuning to the cold cavity. Grey dots: pre-flight. Orange dots: post-flight. Inset: Reverse biased photothermal nonlinear resonance detuning for pre- and post-flight MRR with *p-n* junctions (MRM). Both devices have a $Q_t$ of 22k. Pre-flight samples have higher nonlinear resonance shifts compared to post-flight samples. (**C**) The atomistic picture of highly energetic protons (>10MeV/particle) in space triggered ionization of bounded electrons and thus the mid-gap defects, and (**D**) low energy proton-induced atomic displacement, creating carrier scatter centers and reduce carrier mobility.

Supplementary file for Space Qualifying Silicon Photonic Modulators and Circuits

## S1. Models for the Optical Transmission Spectra

### S1.1 Model analysis of Mach-Zehnder Modulator (MZM)

The lineshapes of the transmission spectra of MZM were fitted by the following equations, to extract the complex effective refractive index ($n_{eff}$):

$$T = \left| \frac{1}{1+\sigma} \left[ \sigma \exp\left( j \frac{2\pi}{\lambda_0} n_{eff}(\Delta L + L) \right) + \exp\left( j \frac{2\pi}{\lambda_0} n_{eff} L \right) \right] \right|^2 \tag{S-1}$$

$\sigma$ is the power splitting ratio of the power splitter, which is approximately 1 for the pre-flight device, given the >25 dB extinction ratio (ER) (grey curves in Fig. 1f and Fig. 2c, e). $L$ is the length of the shorter arm, and $\Delta L = 42\mu m$ is the incremental length of the other arm. $n_{eff}$ is the effective index of the single-mode silicon waveguide (WG). After LEO exposure, the effective index changed to $n_{eff\_post} = n_{eff\_pre} + \Delta n + j\Delta k$, where $\Delta n$ and $\Delta k$ are the radiation-induced change of real and imaginary parts of the effective index. The propagation loss from the material absorption can be correlated to the extinction coefficient ($\Delta k$): $\alpha = 4\pi \Delta k/\lambda_0$. Nearly 10dB reduction of ER (red curves in Fig. 1f and Fig. 2c, e) is attributed to the additional absorption in active WGs.

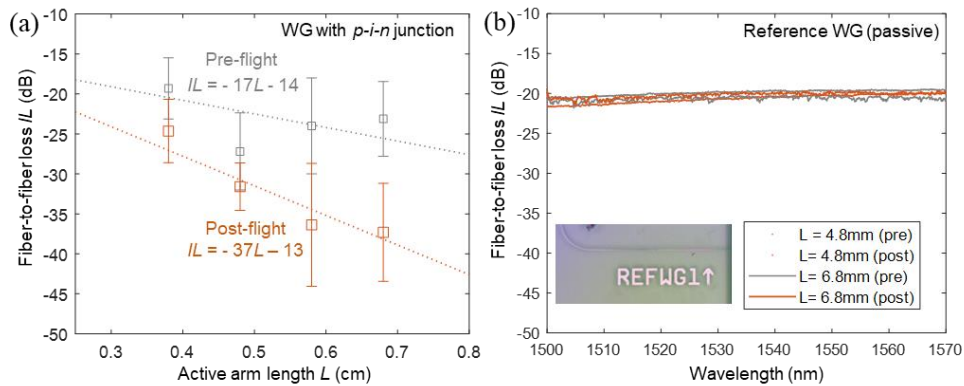

**Fig. S1 | Extraction of propagation loss in MZM with varying arm lengths.** (a) Fiber-to-fiber insertion loss ($IL$) versus the WG length ($L$) with *p-i-n* junction (>18 devices). (b) Transmission spectra for reference WG defined adjacent to those WG with junctions. Passive WGs do not exhibit noticeable $IL$ differences after flight for $L = 0.48$cm and $0.68$cm. Bottom inset: optical microscope image of the inverse taper coupler for the reference WG.

This additional absorption is verified by the increased propagation loss (20dB/cm) of the active arm length (Fig. S1). The fitted propagation loss increases from 17dB/cm (pre-flight, grey) to 37dB/cm (post-flight orange). The total coupling loss (13-14 dB) includes the ones from the power splitters (3dB) and the fiber-chip coupling loss (5dB per facet).

## S1.2 Model analysis of Microring Modulator (MRM)

The transmission spectra of the MRM can be fit by the coupled mode theory (CMT) (48):

$$T(\omega) = |1 - \frac{1/\tau_c}{-j(\omega-\omega_0)+1/\tau_c+1/\tau_{in}}|^2 \qquad (S-2)$$

where $\omega_0$ is the resonance angular frequency of the MRR and $\omega$ is the angular frequency of the input laser. The total cavity lifetime ($\tau_t$) depends on the coupling ($\tau_c$) and intrinsic cavity lifetimes ($\tau_{in}$): $1/\tau_t = 1/\tau_c + 1/\tau_{in}$. $1/\tau_c$ is the loss rates coupled into WG and $1/\tau_{in} = 1/\tau_v + 1/\tau_{lin}$ is the linear loss rate to material absorption ($1/\tau_{lin}$), scattering and radiation ($1/\tau_v$). The relationship between the lifetime and correspondent quality factor is defined as $1/\tau_{c/in/t} = \omega/Q_{c/in/t}$. Given the fixed $Q_c$ for each geometry design, the extinction ratio (transmission $T$ at $\omega_0 = \omega$) depends on $Q_{in}$: ER (dB) = $20 \times [\lg(Q_{in})-\lg(Q_t)]$ (derived from ref. 49), where $1/Q_t = 1/Q_c + 1/Q_{in}$. ER is maximized at $Q_c = Q_{in}$. The propagation loss of the WG can also be easily derived from $Q_{in}$: $\alpha = 2\pi n_{eff}/(Q_{in}\lambda_0) = \lambda_0/R/FSR$ (50-51). The estimated propagation loss of WG w/ $p$-$n$ junction increases from 19.5 dB/cm to 27.4 dB/cm based on the reduced $Q_{in}$ in MRM. R is the radius of the resonator and FSR is the free spectral range on the MRR transmission spectra. The propagation loss from the material absorption can be correlated to the extinction coefficient ($\Delta k$): $\alpha = 4\pi\Delta k/\lambda_0$, which is proportional to $1/\tau_{lin}$.

**Table S-I: Impacts on optical properties of silicon photonic devices**

| Parameters | Dimension | Pre-flight | Post-flight | Difference |
|---|---|---|---|---|
| Propagation loss of passive Si WG | A few cm long | ~3dB/cm | <5dB/cm | Small |
| Propagation loss of active WGs w/ lateral $p$-$i$-$n$ junction | A few cm long | 17dB/cm | 37dB/cm | +20dB/cm |
| $\Delta n_{eff}$ extracted from MZM spectra w/ lateral $p$-$n$ junction* | | -- | -- | $(1+i) \times 10^{-3}$ |
| $Q_{in}$ of passive Si MRR | $R = 20\mu m$ | 130,000 | 120,000 | Small |
| $Q_{in}$ of active MRM w/ $p$-$n$ junction | $R = 10\mu m$ | 35,000 | 25,000 | Reduced |

WG: waveguide; MRR: microring resonator; * Obtained by fitting equation S-1 to the measured spectra (statistics shown in Fig. 2d-e); ** Estimated from the $Q_{in}$ of active MRM.

## S2. Micro-Raman Examination

Fig. S2a shows the Stokes component of the reflection micro-Raman spectra of the 500nm wide passive silicon WG, at the excitation laser wavelength of 532 nm. The peak near 521 cm⁻¹ is crystalline silicon's main transverse optic (TO) phononic mode. Through fitting the experimental data to the Gaussian model, peak frequency, and full wave half maximum (FWHM) of the measured Raman TO peaks are extracted for passive WGs (Fig. S2a) and active WGs with doping (Fig. S2b). Their statistical responses show the peak position remains nearly invariant after LEO exposure (Fig. S2c). For passive intrinsic silicon photonics devices from AIM, the FWHM expands from 3.7 cm⁻¹ (typical for single crystalline silicon) to nearly 4.4 cm⁻¹ (Fig. S2e) (52), while the expansion is not observed in the doped (with *p-n* junction) and undoped WGs from IME (Fig. S2f).

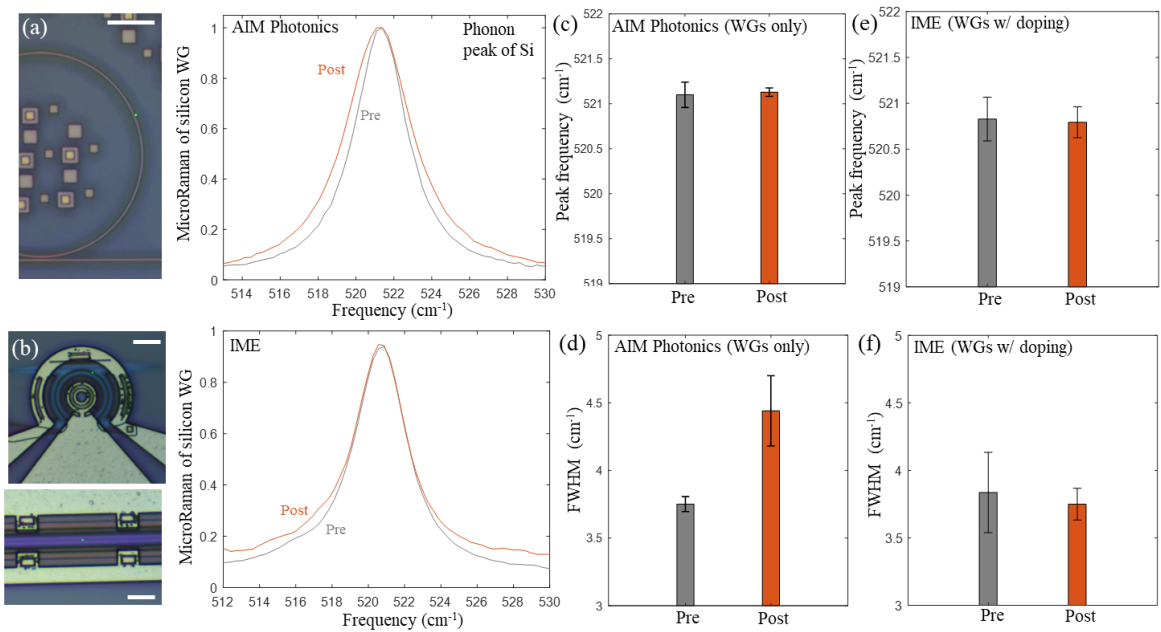

**Fig. S2 | Micro-Raman spectra of the silicon WG pre- and post-LEO exposure.** (a) Micro-Raman spectra near the optical vibrational modes of the silicon WG from different foundries. Optical microscope image of the probe laser on devices from AIM Photonics Foundry (left inset). (b) Micro-Raman spectra near the WG region for modulators from IME Foundry (left insects). (c) Statistical response of peak frequency for AIM and (d) IME devices. (e) FWHM of AIM and (f) IME devices. Grey: control sample. Orange: post-LEO exposure.

It is noted that the intrinsic region in modulators is smaller (<100nm) compared to the vertically incident probing laser spot size (~500nm), the Micro Raman spectra do not reflect the clean signal from the intrinsic region of active WGs, but primarily of the doped region. The same trend is found for an adjacent reference device close to the MZM without doping. The expansion of the FWHM is associated with the degraded crystal symmetry in single crystalline silicon, which is attributed to particle radiation damage. Such change is only observed in passive WG from AIM photonics. The undoped WG is not affected after LEO exposure. Other weak micro-Raman peaks, such as the second-order peak near 950–100 cm$^{-1}$ remain invariant after LEO exposure.

## S3. Impacts of Heavy Ions

Among all the MZMs with long arms, two adjacent pairs of MZMs failed (marked in Fig. 2b). No interference patterns are observed on the transmission spectra, which means one of the active arms is disconnected (Fig. S3). The characteristic of low possibility but disruptive damage to the silicon nanowire aligns with the behavior of heavy ions from GCR, including low flux ~4 particles cm$^{-2}$s$^{-1}$, high atomic number, and high energy (15). Deposition of a single heavy ion from GCR on a silicon WG leads to a disconnected cm-long active arm in MZM.

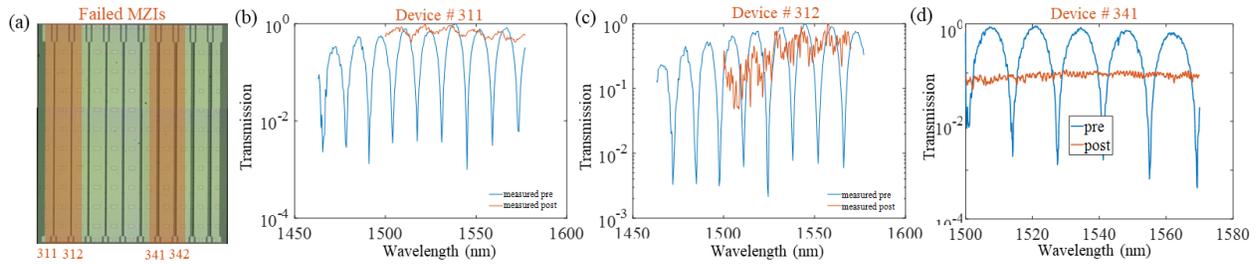

**Fig. S3 | Disrupt damage of heavy ions on MZMs with cm-long active arm.** (a) Top microscope view of the MPW die with failed MZIs marked on the chip (second row of devices in Fig. 2a). (b) Measured transmission spectra of device 311 before (blue) and after the flight (orange). (c) results for devices 312 and (d) 341.

## S4. DC Electronic Response

For both *p-n* and *p-i-n* junctions, the input voltage (*V*) dependent readout current (*I*) complies with the empirical diode equation (40):

$$I = I_0 \left[ \exp\left(\frac{V - I*R_s}{nV_T}\right) - 1 \right] + \frac{V}{R_{SH}} \qquad \text{(S-3)}$$

Where $I_0$ is the reverse saturation current, $R_{s/SH}$ is the series/shunt resistance, $n$ is the ideality factor, $V_T$ is the thermal voltage at room temperature (25.9 mV). The reverse saturation current is inversely related to the square root of the minority carrier lifetime ($\tau_{rec}$). The reduced $\tau_{rec}$ increases the contribution from the recombination current and thus increases the ideality factor ($n$). A summary of those parameters impacted by the collective ionizing radiation is listed in Table S1. Reverse saturation current ($I_0$) and shunt resistance ($R_{SH}$) are extracted by fitting the reverse biased range of the dark IV curve. The ideality factor n indicates the contributions from diffusion and recombination to the total current (53).

## S5. High-speed Optoelectronic Bandwidth for Integrated Photonic Modulators

### S5.1. Optoelectronic Bandwidth for MZM (cm length)

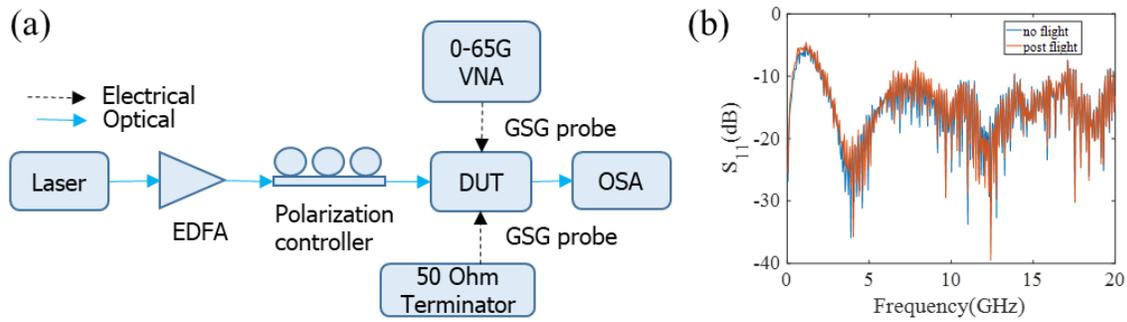

**Fig. S4 | High-speed optoelectronic test setup schematics.** (a) Details of the optical spectra analysis method for MZM. (b) Simultaneously captured S11 for the RF transmission line of MZM (without optical excitation).

The optoelectronic bandwidth of the MZM is limited by the carrier response time ($\tau_e$) and RC constant ($\tau_{RC}$):

$$\frac{1}{f_{3dB}} = 2\pi(\tau_e + \tau_{RC}) \qquad \text{(S-4)}$$

The carrier response time depends on the carrier transient time ($\tau_{tr}$) and $\tau_{rec}$ : $\frac{1}{\tau_e} = \frac{1}{\tau_{rec}} + \frac{1}{\tau_{tr(V)}}$, where the carrier transient time $\tau_{tr}$ is reverse bias (V) dependent, and the minority carrier recombination lifetime $\tau_{rec}$ = 500 ps can be reduced by doping and radiation (in a non-intentionally doped silicon device layer). The RC constant at low frequency can be estimated by comparing the products of resistance and capacitance (Fig. 3c). The RC constant variation can be estimated by comparing the S11 response (54). The overlapping S11 spectra for pre- and post-flight samples indicate a very limited change in RC constant (Fig. S4).

## S5.2 Optoelectronic Bandwidth for MRM (μm size)

The 3dB modulation bandwidth of MRM is limited by the cavity photon lifetime ($\tau_t$), in addition to the electronic design limited response time as MZM:

$$\frac{1}{f_{3dB}^2} = (2\pi\tau_t)^2 + [2\pi(\tau_e + \tau_{RC})]^2 \tag{S-5}$$

Where $\tau_t$ can be derived from the total quality factor of the microring: $\tau_t = Q_t/\omega_0$. Given the extracted total quality factor ($Q_t \sim 15k$). $\tau_t$ is estimated to be around 5 ps, which is much smaller than the electronically limited lifetime.

## S6. Free-carrier Lifetime Dependent Nonlinear Photo-thermal Effect in Silicon Microring

At increasing input optical power, the photon excited free carrier and thermal effects impact the light interaction with the MRR. The nonlinear CMT correlating the dynamics of photon, electron, and temperature is (41, 51):

$$\frac{da}{dt} = \left(i(\omega_L - \omega_0 + \Delta\omega) - \frac{1}{2\tau_t}\right)a + \kappa\sqrt{P_{in}} \tag{S-6}$$

$$\frac{dN}{dt} = \frac{1}{2\hbar\omega_0\tau_{TPA}}\frac{V_{TPA}}{V_{FCA}^2}|a|^4 - N\left(\frac{1}{\tau_{rec}} + \frac{1}{\tau_{tr}(V)}\right) \tag{S-7}$$

$$\frac{d\Delta T}{dt} = \frac{R_{th}}{\tau_{th}}\left(\frac{1}{\tau_{FCA}} + \frac{1}{\tau_{lin}}\right)|a|^2 + \frac{\Delta T}{\tau_{th}} \tag{S-8}$$

Where $a$ is the amplitude of resonance mode; $N$ is the free-carrier density; $\Delta T$ is the cavity temperature shift. $P_{in}$ is the power carried by an incident continuous-wave laser. $\kappa = \sqrt{\frac{1}{\tau_c}}$ is the coupling coefficient between the WG and cavity, where $\tau_c$ is the coupling limited lifetime between the bus WG and MRR. $\omega_L$-$\omega_0$ is the detuning between the laser frequency ($\omega_L$) and cold cavity resonance ($\omega_0$). The total cavity resonance shift is $\Delta\omega = \Delta\omega_N - \Delta\omega_T$, where $\Delta\omega_T$ is the thermal dispersion and $\Delta\omega_N$ is the free-carrier dispersion. The total photon loss rate is $1/\tau_t = 1/\tau_c + 1/\tau_v + 1/\tau_{lin} + 1/\tau_{TPA} + 1/\tau_{FCA}$. The linear loss rate $1/\tau_{lin}$ represents the linear material absorption rate by the mid-gap defect states.

$1/\tau_{FCA}$ is the free-carrier absorption rate. $1/\tau_{TPA}$ is the two-photon absorption rate. $V_{FCA}$ is the effective mode volume for free carriers and $V_{TPA}$ is the effective mode volume for two-photon absorption. In steady-state conditions, the photon amplitude at relatively low input power ($P_{in}$) can be approximated as $a = 2k\tau_e\sqrt{P_{in}}$. The relationship needs to be rewritten at high input power as $a \propto \tau_e^{-1/5}P_{in}^{1/10}$. Within the low nonlinear intensity region, the input power-dependent carrier density is $N \propto \tau_e P_{in}^2$ (derived from equation S-7), and the corresponding incremental temperature is $\Delta T \propto \tau_e P_{in}^3$ (derived from equation S-8). We know that the resonance shift ($\Delta\lambda_0$) is thermally induced ($\Delta T$): $\Delta\lambda_0 \propto \Delta T \propto \tau_e P_{in}^3$ is the relation used to calculate the change of the free-carrier lifetime. Given the free-carrier lifetime (500ps) of silicon WG (51), we derived the post-flight free-carrier lifetime of 226.16 ps, by comparing the nonlinear optical resonance shift of MRR with similar quality factor and the same excitation power level. This model is used to interpret the nonlinear transmission spectra in Fig. 4. Note that the total carrier lifetime $\tau_e$ is affected by carrier transient time ($\tau_{tr}$) and radiation-dependent recombination lifetime ($\tau_{rec}$) (equation S-4).

Then we proceed to examine the origin of the radiation leading to significantly reduced $\tau_e$. To differentiate between TID and DDD, we bake the sample at increasing temperatures (up to 300ºC) and examine the nonlinear response (Fig. S5). The carrier recombination lifetime seems not recovered after annealing, identifying the source of DDD leading to the reduced $\tau_e$. A higher annealing temperature (typically 600ºC) can heal the dangling bonds (and thus recover $\tau_{rec}$), but also cause permanent damage to doping profiles and electrodes.

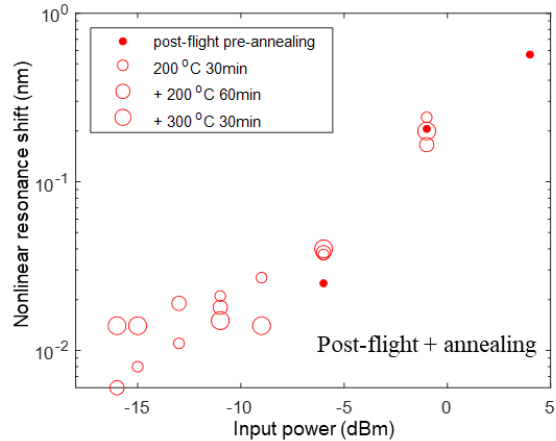

**Fig. S5 | Nonlinear resonance shift of the post-flight MRR after accumulative baking.**

## S7. Summary of the Optoelectronic Response Change after LEO Exposure

The parameters in Tables S-I and S-II are the results from the nanoscale optoelectronic device comparison (same device or representative device in the same MPW run with the same design).

**Table S-II: Impacts on the optoelectronic parameters**

| Parameters | Source | Pre-flight | Post-flight | Difference |
|---|---|---|---|---|
| Reverse saturation current/leakage current, $I_0$ (pA) | Dark IV | 118 | 256 | ++ |
| Series resistance, $R_s$ ($\Omega$) | Dark IV | 133* | 131* | Trivial |
| Carrier mobility | $\sim 1/R_s$ | | | Trivial |
| Shunt resistance, $R_{SH}$ (T$\Omega$) | Dark IV | 18.6 | 16.6 | Trivial |
| Ideality factor | Dark IV | 1.39 (p-n) | 1.44 (p-n) | Slight increase |
| | | 1.47 (p-i-n) | 1.51 (p-i-n) | |
| Capacitance, $C$ (fF) | C-V | 72.2 | 78.1 | Trivial |
| RC constant, $\tau_{RC}$ (ps) | IV & CV | 8.46* | 7.54* | Trivial |
| Carrier recombination lifetime (or free carrier lifetime) $\tau_{rec}$ (ps) | Nonlinearity in microring | 500 (53) | 226 | <1/2 |
| Optoelectronic bandwidth, $f_{3dB}$ (GHz) | High-speed modulation | ~5 | ~10 | ×2 |

* The real values of series resistance and RC constant are likely to be smaller than the estimated values here (extracted from fitting the IV curves upto 1.5V bias).

## S8. Comparison to Ground Tests and Defect Type Analysis

Compared to the 18 experiments of the ground radiation tests that we can find (summarized in table S-III), we provided the most comprehensive and systematic studies of those nanoscale optoelectronic devices and circuits, trying to capture the full spectrum of cosmic radiation impacts on nanostructured optoelectronic devices, as well as tracking the atomic origin of the defects, as conventional spectroscopy tools only apply on uniform and low defects single crystalline films. The characterizations are summarized as 9 aspects: (1) utilizing the micro-spectroscopy tool to track the atomic origin of the defects (Section S2); (2) characterized propagation loss through length varying devices (Section S1), (3) effective index variations (Table S-I); (4) revealed doping dependence of radiation impacts by comparing passive device and doped devices with the identical nanostructures (Section S1 and Table S-I); (5) electro-optic tuning efficiency (Fig. 2, Table S-II); (6) high speed optoelectronic bandwidth (Fig. 3 and Table S-II); (7) detailed electronic characterization and diode analysis by comparing nanostructured *p-n* and *p-i-n* junctions (Fig. 2 and Table S-II); (8) nonlinear optic response of the micro-cavities (Fig. 4) and (9) the influence of the device footprint (from ~10 μm to mm size, Fig. 2).

**Table S-III | Comparison of the ground and space radiation test conditions**

| Ref. | Radiation | Dosage | Energy/Particle | Material/Device | Radiation impacts | DC EO | RF EO |
|---|---|---|---|---|---|---|---|
| 19 | γ | 0.1Mrad | $10^6$ eV | Si MRR | TID ($\Delta n_{eff}\downarrow$) | - | - |
| 20 | γ | 0.1Mrad | $10^6$ eV | Si MZI | Propagation loss ↑ | - | - |
| *55* | γ | 0.3 Mrad | $10^6$ eV | InGaAsP/InP MRR | TID, $\Delta n_{eff}\uparrow$ $Q\downarrow$ (-30%) ER ↓ | - | - |
| *56* | γ | 1-15Mrad | $10^6$ eV | a-Si MRR | TID | - | - |
| *57* | γ | 10Mrad | $10^6$ eV | a-Si, $SiN_x$ MRR | TID | - | - |
| *58* | γ | 10Mrad | $10^6$ eV | SiC MRR | TID, $\Delta n_{eff}\uparrow$ | - | - |
| 21 | γ | 40Mrad | $10^6$ eV | Si MRR, MZI | TID, $\Delta n_{eff}\downarrow$ | - | Error rate↑ |
| 22 | γ | 100Mrad | $10^6$ eV | Si MRR | TID | - | - |
| 19 | X-ray | 6.7Mrad | $10^4$ eV | Si MRR** | TID, $\Delta n_{eff}\downarrow$ | - | - |
| 26 | X-ray | $10^3$Mrad | $10^4$ eV | Si MRM, MZM (Doped) | TID | Reduced EO tuning | - |
| 23 | X-ray | 100Mrad | $10^4$ eV | Si MZM (Doped) | TID, $\Delta n_{eff}\downarrow$ | Reduced EO tuning | - |
| 24 | X-ray | >$10^6$Mrad | $10^4$ eV | Si MZI | TID, $\Delta n_{eff}\downarrow$ | -- | - |
| 25 | X-ray | >$10^6$Mrad | $10^4$ eV | Si MRM (Doped) | TID | Reduced EO tuning | - |
| 59 | Neutron | 2.5Mrad | 2 $10^7$ eV | Si MRM (Doped) | No change | No change | - |
| 23 | Neutron | 100Mrad | $10^7$ eV | Si MZM (Doped) | No change | No change | - |
| 59 | Neutron | $10^{12}$ /cm² (high) | $10^7$ eV | Si MRR, MZI | No change | N/A | - |
| *30* | Proton | 100Mrad | 4sources | SiN MRR | No change | - | - |
| *60* | α | $10^{15}$ cm$^{-2}$ | $10^6$ eV | SiON MRR | $\Delta n_{eff}\uparrow$, $Q\downarrow$ (-20%) | - | - |
| This work | LEO (X-, γ-ray, proton, heavy ions) | 14.85 rad† | Upto $10^7$- $10^{11}$ eV | Si MRM, MZM (Doped) | Reduced ER (-10dB) and $Q$ (-30%) | No change | Faster |
| | | | | Si MRR, MZI | | Reduced thermal nonlinearity | - |

*Tilted* references: not c-Si photonic integrated circuits

† Includes both cosmic ray and particle radiations absorbed by silicon-based dosimeter

Those reported ground tests are significantly different from the radiation background on LEO: (1) heavily overdosed compared to LEO space environment. (2) focusing on a single energy level γ or X-ray, while the space cosmic rays cover a broad range of spectra. A similar case is applied to X-ray. (3) Critical particle radiation is much less studied, limited to a few studies of neutron exposure or on passive device studies with negative results. (4) a broad range of energetic protons exposure on actives is missing, (5) limited to DC optoelectronic characterizations (Fig. S6a).

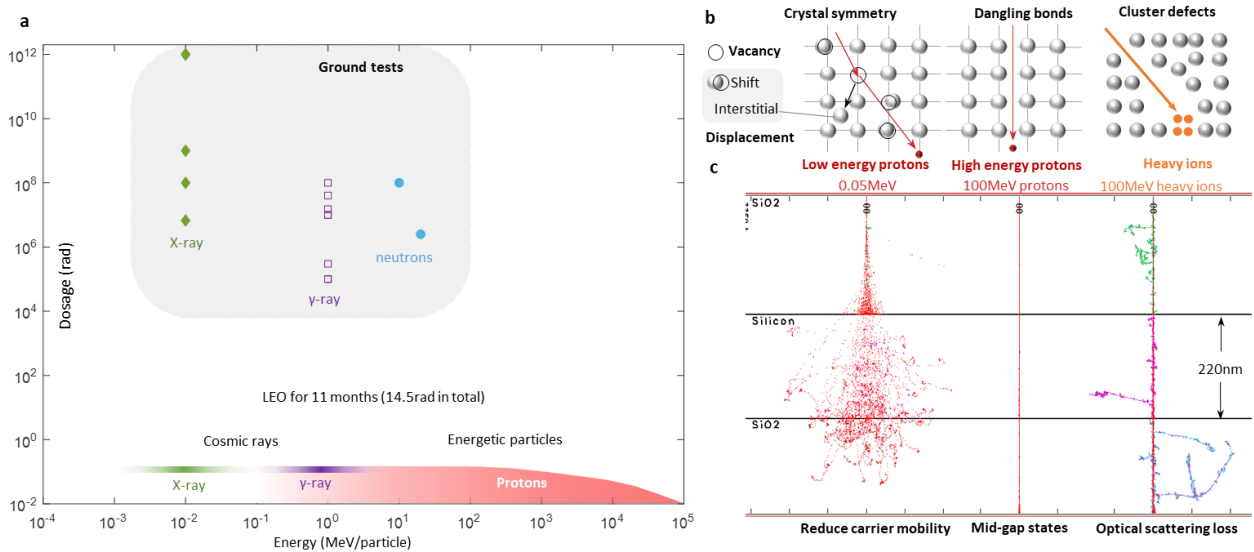

**Fig. S6 | Comparison between ground tests and year-long radiation on LEO.** (a) Radiation environment comparison between ground tests reported so far and the real space environment (Table S-III). (b) Atomistic pictures of the particle radiation-induced defects in silicon crystalline lattice, for low energy/speed protons produced vacancy and displacement (left), high energy protons induced dangling bonds (middle) and heavy ions resulted in cluster defects (right). (c) Simulated particle trajectory in silicon-on-insulator substrate (220 nm silicon layer), for low-energy protons (left), high-energy protons (middle), and high-energy heavy ions (right). The left one is more available on ground test, while high energy protons and heavy ions are more prevalent in space.

The atomic origin of the particle radiation damage is associated with the creation of defects in single-crystalline lattices, which can be in the form of a change in the crystalline symmetry, formation of silicon vacancies, and interstitial atoms. For low-energy proton exposure, more vacancies and interstitial atoms are created near the device layer, creating carrier scatter centers, and reducing carrier mobility. After cosmic radiation exposure with low density and high energy proton, the resistance and thus carrier mobility is not changed, but we notice the creation of mid-gap states as carrier recombination centers (increase ideality factor and reduce free carrier lifetime). The cluster defects are more likely to be induced by heavy ions/nuclei, which is the only mechanism that destroys nanostructures and disrupts the waveguide propagation (Fig. S6b-c). It is noted that due to the nanoscale device statistics, we cross-compare multiple characterization results/multiple devices to identify the nanostructured material response (summarized in Table S-IV).

**Table S-IV | Radiation induced defect types and silicon photonic device response**

|  | Ionizing radiation (X, γ-rays) | | Particle radiation (Protons, α, neutrons) | | Heavy ions |
|---|---|---|---|---|---|
| Atomic origin | Ground: Mrad | **LEO:** 15 rad | Ground: low energy, high dose | **LEO:** High energy charged particles | **Space only** |
| Surface oxidation | $\Delta n_{eff}\downarrow$ (2,3) | -- | -- | -- | -- |
| Crystal asymmetry (1) | -- | -- | $\Delta n_{eff}\uparrow$, Scattering loss in WG (2,3) | $\Delta n_{eff}\uparrow$ Doping dependent $\Delta k \uparrow$ (2-4) | -- |
| Mid-gap states | Reduced EO tuning (5) | -- | -- | Reduce carrier lifetime (6,7,8) | -- |
| Custer defects | -- | -- | -- | -- | High loss (9) |

[*] Observations: (1) Micro-spectroscopy, (2) propagation loss (or $Q$), (3) $\Delta n_{eff}$, (4) doping dependent radiation damage, (5) EO tuning, (6) high speed EO, (7) diode response, (8) optical nonlinearity, (9) MZM with long arm of doped waveguides.

Here we just focus on silicon photonics (single crystalline silicon on insulator substrate). Amorphous materials (such as SiNx, a-Si) are not sensitive to radiation damage [even with upto $10^7$ rad γ ray (58), or $10^{15}$ cm$^{-2}$ α particles (60)] given their highly defective nature. Micro-Raman spectroscopy in section S2 shows broadened bandwidth of the silicon waveguide (device layer embedded under oxide with metal interconnects), suggesting that the crystalline symmetry degradation in some passive waveguide, which is likely to be caused by the high energy charged particles (protons) given such low accumulative dosage (Table S-IV).